\documentclass{aa}
\usepackage{graphicx,natbib}
\usepackage{txfonts}
\begin{document}
   \title{Large-scale variations of the dust optical properties in the
   Galaxy}

   \author{L. Cambr\'esy\inst{1}
          \and
          T.H. Jarrett\inst{2}
	  \and
	  C.A. Beichman\inst{3}
          }

   \offprints{L. Cambr\'esy}

   \institute{Observatoire Astronomique de Strasbourg, F-67000 Strasbourg,
		France\\
		\email{cambresy@astro.u-strasbg.fr}
	\and
		Infrared Processing and Analysis Center, 100-22, California
		Institute of Technology, CA 91125\\
		\email{jarrett@ipac.caltech.edu}
	\and
		Michelson Science Center, Infrared Processing and Analysis
		Center, 100-22, California
		Institute of Technology, CA 91125\\
		\email{chas@ipac.caltech.edu}
}

   \date{Received ; accepted }

   \abstract{We present an analysis of the dust optical properties at large
   scale, for the whole galactic anticenter hemisphere. We used the 2MASS
   Extended Source Catalog to obtain the total reddening on each galaxy line
   of sight and we compared this value to the IRAS 100 $\mu$m surface
   brightness converted to extinction by \citet{SFD98}. We performed a 
   careful examination and correction of the possible systematic effects
   resulting from foreground star contamination, redshift contribution and
   galaxy selection bias. We also evaluated the contribution of dust 
   temperature variations and interstellar clumpiness to our method.
   The correlation of the near--infrared extinction to the far--infrared
   optical depth shows a discrepancy for visual extinction greater than 1 mag
   with a ratio $A_V({\rm FIR}) / A_V({\rm gal}) = 1.31 \pm 0.06$. We attribute
   this result to the presence of fluffy/composite grains characterized by an
   enhanced far--infrared emissivity. Our analysis, applied to half of
   the sky, provides new insights on the dust grains nature suggesting fluffy
   grains are found not only in some very specific regions but in all
   directions for which the visual extinction reaches about 1 mag.
   }

   \maketitle
%
%________________________________________________________________

\section{Introduction}
Large scale dust studies started with IRAS which revealed the whole sky in
four far--infrared (FIR) wavelength bands. It led to the current view of the
interstellar dust seen as a 3 components grain population \citep{DBP90}:
polycyclic aromatic hydrocarbons (PAH), very small grains (VSG) and big
grains (BG). The color analysis of IRAS surface brightnesses provides insight
to the variations of the 3 dust component abundances with
respect to various environments. BG dominate in molecular clouds whereas VSG
are more representative of the diffuse interstellar medium usually associated
with the atomic hydrogen.
Our understanding of interstellar medium (ISM) dust has further evolved with
advances in FIR and submillimeter (submm) instrumentation in the last decade.
Not only is the dust composed of different grain size components but the
optical properties of the dust change with the environment. 
The PRONAOS balloon observations reveal the complexity of the dust
properties with emissivity variations \citep{BAR+99,SAB+03a} and spectral
index variations \citep{DGB+02}. More difficult because of the atmosphere
absorption but still possible, ground based observations in the
submillimeter range also find evidence of dust emissivity variations
with SCUBA observations \citep{JFR+03}.
The initial results from IRAS, DIRBE, or ISO, which were mainly focused on
dust abundances and temperature, reveal much more if re-examined in light
of these recent dust optical property results obtained from longer
wavelengths, notably in the submm. 
\citet{CBLS01} have quantified the emissivity enhancement at
100 $\mu$m comparing COBE/DIRBE with an extinction map of the Polaris Flare.
Their results have been confirmed by \citet{dLAK03} who found similar
behavior for the FIR dust emissivity in eight nearby interstellar regions
observed by ISOPHOT.
\citet{DH02} studied a sample of galaxy spectral energy distributions with
ISO and IRAS, proposing a change in the dust emissivity at wavelengths longer
than 100 $\mu$m to match SCUBA brightnesses. ISOPHOT sources discovered at
200 $\mu$m (but not detected at 100 $\mu$m) suggest a change in the
grain properties \citep{LML+03}.\\ 

Although it is now established that the dust optical properties vary with
the environment, likely due to fluffy grain population, the spatial extent
in these variations is still controversial and the density threshold for the
transition from regular to the enhanced emissivity dust population remains
unknown.
In this paper we address these questions using a large scale analysis of
the whole galactic anticenter hemisphere. 
We propose to compare data from IRAS and DIRBE, converted to dust extinction
by \citet{SFD98} (hereafter SFD98), with the reddening of 2MASS galaxies. Our
goal is to investigate the dust properties through the apparent discrepancy
between these two quantities.
The 2MASS extended source catalog characteristics are presented 
in Sect. \ref{2masscat} followed by a detailed analysis of the biases in the
galaxy color excesses in Sect. \ref{bias}. Section \ref{results} is dedicated
to the analysis of the correlation between the galaxy reddening and the FIR
dust opacity, the conclusions of the paper are presented in
Sect. \ref{conclusion}.

\section{The 2MASS extended source catalog}\label{2masscat}
\subsection{Presentation}
The 2MASS All-Sky Data Release for extended sources (XSC) contains positions
and photometry in $JHK_s$ for 1.6 million objects \citep{JCC+00b}. About 97\%
are galaxies but the catalog also contains galactic resolved sources such
as HII regions, planetary nebulae, reflection nebulae, young stellar objects,
etc. Galaxies were identified using a {\it decision tree} method which uses
several parameters including size, shape, central surface brightness and color.
In order to increase the reliability of the object classification, a {\it
visual classification} has also been performed. It is mainly based on the eye
examination of the color-combined images, but also optical DSS images when it
was needed. Roughly 25-30\% of all XSC sources have been inspected, with close
to 100\% for bright sources \citep[$K_s<12.5$ mag,][]{CSv+03}. All XSC sources
have been inspected in the galactic plane
($|b|<5\deg$) and in some other regions of specific interest (Magellanic
clouds, galaxy clusters like Virgo). Visual inspection reveals that only a
small fraction (less than 1\%) of sources are artifacts of various origins
such as bright star neighborhood, airglow, meteor streaks or image edges.
In total, among the $\sim 1.5 \times 10^6$ galaxies in the XSC, more than
$3 \times 10^5$ have been confirmed by eye.

\subsection{Catalog properties}
\subsubsection{Source density and completeness}
\begin{figure*}
	\centering
	\includegraphics[width=18cm]{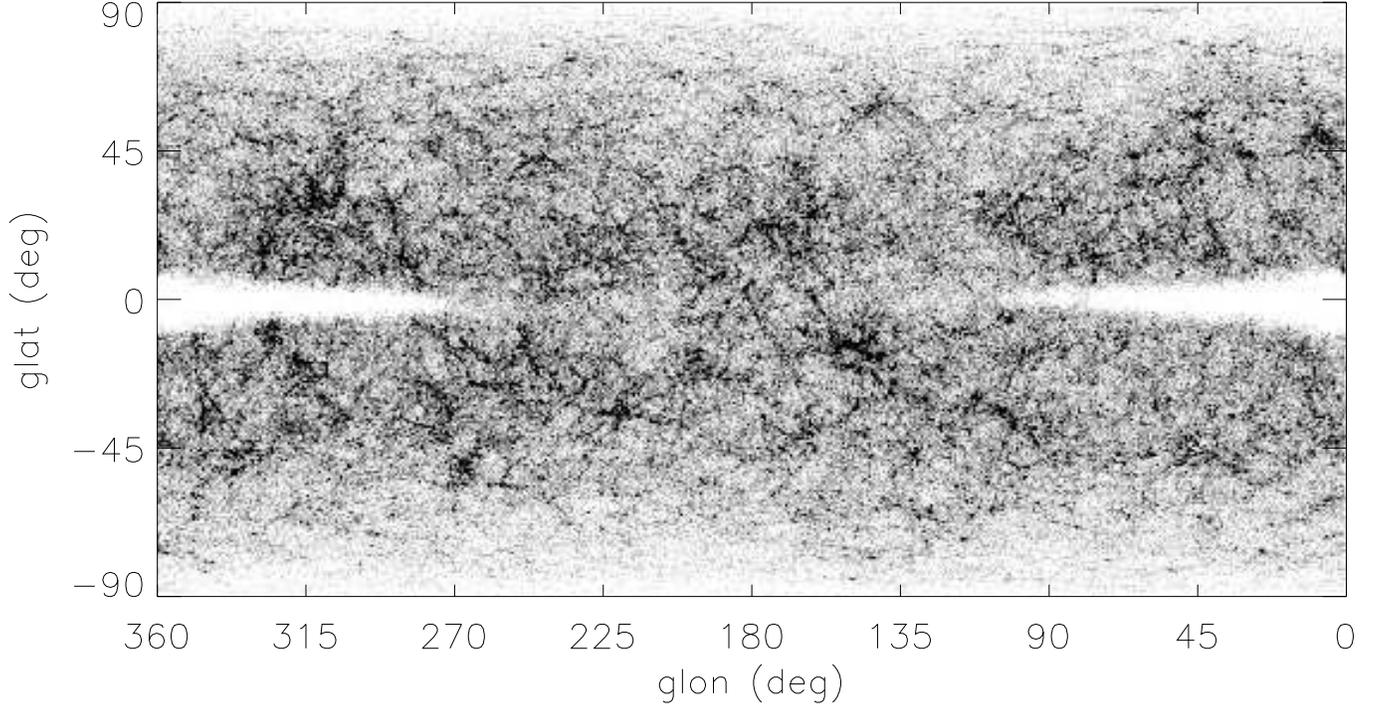}
	\caption{Spatial distribution of the $1.5 \times 10^6$ 2MASS galaxies
	projected in rectangular coordinates. Milky Way objects have been
	removed. The Cartesian projection is in galactic coordinates and is
	centered at the anticenter direction. The absence of detected galaxies
	near the galactic center results from the confusion noise due to
	high stellar densities.}
	\label{galdistrib}
\end{figure*}

Figure \ref{galdistrib} shows the spatial distribution of the XSC galaxies,
filtered from the Milky Way extended sources. The large scale structures of
the universe dominate the source distribution over the local dust extinction.
The confusion noise prevents the galaxy identification in the very high
stellar density regions such as the Galactic Center \citep{JCC+00a}.
The striking filamentary structure in the source distribution reflects the
large scale structure of galaxies in the universe \citep[see also][]{Jar04}.
It is worth noting that if Fig. \ref{galdistrib} were restricted to a
longitude range of 90 to 270 deg (i.e. the anticenter hemisphere), the Milky
Way disk would barely be seen. We will therefore not consider using source
counts in the near--infrared as an extinction estimator as it has been done by
\citet{BH82} in the visible. Since the source counts are incomplete in the
Galactic Center region, the analysis presented in this paper is restricted
to the anticenter hemisphere, about $8.5 \times 10^5$ galaxies.

\begin{figure}
	\centering
	\includegraphics[width=8.0cm]{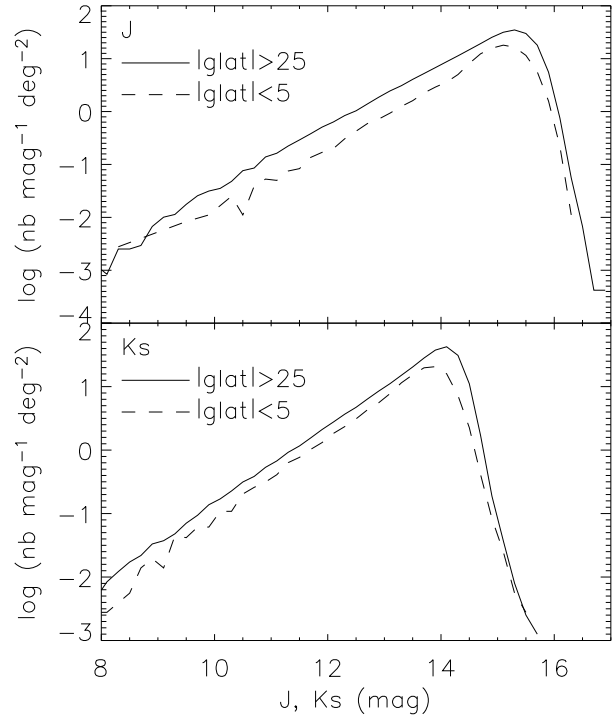}
	\caption{Magnitude distribution of galaxies at $J$ (upper panel)
	and $K_s$ (lower panel) for latitudes greater than 25 deg and lower
	than 5 deg. Only sources in the galactic anticenter hemisphere
	(longitude from 90 to 270 deg) are considered.}
	\label{magdistrib}
\end{figure}
In the following, magnitudes always refer to the fiducial elliptical isophotal
($K_s = 20$ mag arcsec$^{-2}$) magnitudes. The level of completeness for the
detection depends on the latitude as the stellar density increases in the
plane. Magnitude distributions for $|b|>25\deg$ and $|b|<5\deg$ are presented
in Fig. \ref{magdistrib} \citep{JCC+00a}. The low and high latitude samples
are quite similar. The latitude has only a small impact on the value of the
turn off in the distribution which represents the 2MASS sensitivity
limit for extended objects. The low-latitude distribution is almost parallel
to the high-latitude curve, indicating the confusion noise reduces the
completeness at all magnitudes (except the very bright end, $J<9$) without
any significant change in the observed luminosity function shapes.
Consequently, we do not expect any significant color systematic effect
from photometry signal-to-noise.

\subsubsection{Galaxy colors}\label{galcol}
The color distribution of galaxies is related to their morphological
type. It reflects various characteristics such as the stellar
content (old/young populations, star forming rate) and the dust fraction.
\citet{Jar00} and \citet{JCC+03} analyzed 2MASS galaxy properties according
to their morphological types and found that ellipticals are slightly redder in 
near--infrared color than lenticulars and that normal barred and transition 
barred spirals have similar colors. However the dispersion for spirals is
increased by a factor of 2, going from early to late type due to important
star formation and amount of dust material in the earliest types. For some
morphologies, color can be clearly different as for AGN which are 0.3-0.4 mag
redder than the average, or dwarf, irregular/peculiar and compact galaxies
which are bluer than the norm.
Ellipticals and spirals dominate the galaxy population and we find 
$J-K_s = 1.0 \pm 0.1$ for high-latitude galaxies (no Milky Way extinction)
restricted to $J<14.0$ and $K_s<13.3$. These magnitude cuts are addressed 
later in the paper and correspond to our high-latitude sample as shown
Fig. \ref{maglim}.

\section{Extinction from galaxy colors}\label{bias}
The extinction estimation is based on the $J-K_s$ galaxy color excess from
the normal distribution. Visual extinction is derived from the color
excess $E_{J-K_s}$ as follows:
\begin{eqnarray*}
E_{J-K_s} &=& (J-K_s)_{\rm obs}-(J-K_s)_{\rm int}\\
A_V &=& \left(\frac{A_J}{A_V} - \frac{A_{K_s}}{A_V}\right)^{-1} E_{J-K_s}  
\end{eqnarray*}
where the indices $obs$ and $int$ stand for $observed$ and $intrinsic$,
respectively. $(J-K_s)_{\rm int} = 1.0 \pm 0.1$ mag assuming zero redshift.
We use the extinction law from \citet{RL85} for which $A_J/A_V=0.282$ and
$A_{K_s}/A_V=0.112$ yielding the color excess to visual extinction conversion
formula $A_V=5.88 \times E_{J-K_s}$. 
Although the determination of the extinction looks straightforward,
systematic effects might occur and must be overcome. We distinguish three
independent possible effects: 
\begin{itemize}
\item contamination of the photometry by foreground stars
\item color of a galaxy affected by its redshift.
\item selection effect with the galactic latitude.
\end{itemize}
The following section explores all the possible bias that may contaminate the
galaxy colors.

\subsection{Biases in the reddening estimation}\label{bias123}
\subsubsection{Foreground star contamination}
\begin{figure}
	\centering
	\includegraphics[width=8.8cm]{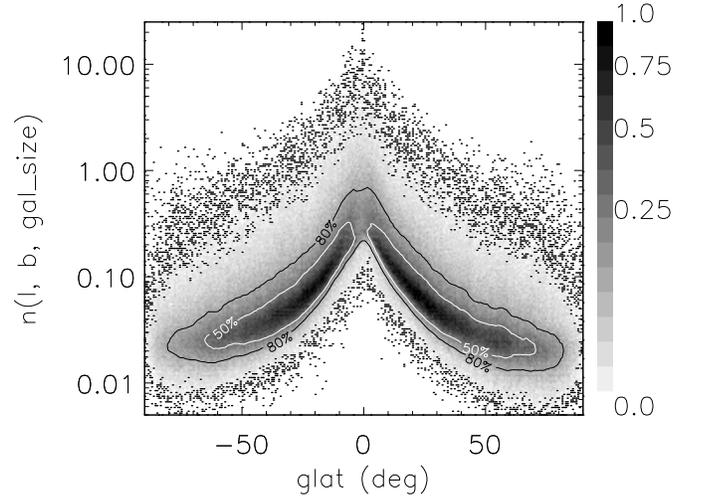}
	\caption{Density plot of the number of foreground stars from a
	Monte Carlo simulation. The contamination probability depends on the
	latitude, the longitude (from 90 to 270 deg) and the galaxy size which
	are taken from the XSC.
	The y-axis is in log scale and the bar scale indicates the normalized
	density color coding. As an example, for a galaxy at $b=10$ deg, the
	number of foreground stars is between 0.04 and 6 (depending on
	longitude and galaxy size) and the most common number is about 0.2
	star (within the 50\% iso-contour).}
	\label{fgstar}
\end{figure}
\begin{figure*}
	\centering
	\includegraphics[width=18cm]{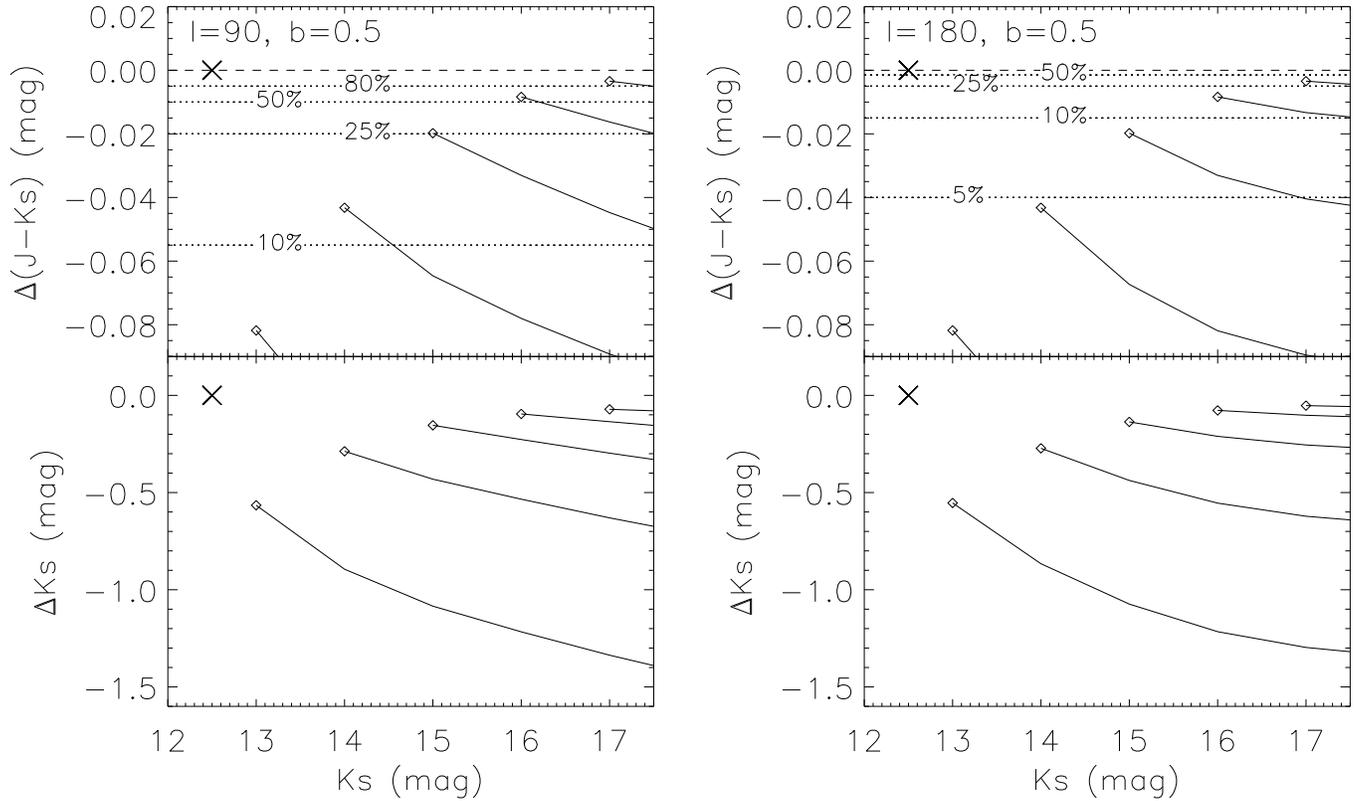}
	\caption{$J-Ks$ color contamination (upper panel) and $K_s$
	contamination (lower panel) of a typical galaxy ($K_s=12.5$, with a
	12\arcsec\ diameter) by foreground stars in the galactic plane for
	$l=90$ deg (left) and $l=180$ deg (right). In all panels, the cross
	represents the reference galaxy and the diamonds the brightest
	foreground star. Solid lines show the contamination level versus the
	magnitude of the foreground stars. For the upper panels, dotted lines
	give the probability of reaching the corresponding contamination
	level.}
	\label{fgstarcontam}
\end{figure*}
The probability of having a star on a galaxy line of sight depends on the
galaxy size and the local stellar number density. The first parameter
introduces a systematic effect because galaxies of a given size are not
randomly spread over the sky but often concentrate in clusters which are
groups of same distance, and thus similar size, objects. The latter parameter,
the stellar number density, relies mostly on the galactic latitude or the
star distribution in the disk of the Milky Way.
The foreground contamination biases the galaxy photometry toward the blue
because stars are generally bluer than galaxies with an average $J-K_s$ color
of $\sim 0.65$ versus $\sim 1.0$ (see Sect. \ref{galcol}).
However, if the star is bright enough compared to the galaxy it is likely
identified and its contribution to the total flux is subtracted
\citep{JCC+00b}. The contamination arises for stars more than 1 magnitude
fainter than the galaxy. A Monte Carlo simulation is well adapted
to estimate the bias. We use the following parameters:
\begin{itemize}
\item $\; \left[J({\rm gal})+1\right] \;   <   J({\rm star}) \; < 19.0$ and 
$\left[K_s({\rm gal})+1\right] < K_s({\rm star}) < 17.5$. Fainter sources
comprise the uniform background level (i.e. background fluctuations).
\item galaxy size and local stellar number density for $K_s<14$ from the XSC
catalog.
\item star luminosity function slope (to extrapolate the number of star
fainter than $K_s=14$): 0.31 
\item $\overline{J-K_s}$ from 0.65 mag for stars at high galactic latitude to
0.98 mag at $b=0$ were diffuse extinction is not negligible and reddening is
appreciable. 
\end{itemize}
Figure \ref{fgstar} shows the number of foreground stars for each galaxy versus
the galactic latitude. Using this number, the stellar luminosity
function and the magnitude limit for stars mentioned previously we removed
the stellar component in the flux of each galaxy. The level of contamination
is globally small since less than 1\% of the galaxies have at least 1
foreground star.
For latitude higher than 25 deg or lower than 5 deg the fraction of
contaminated galaxies is 0.09\% and 1.5\%, respectively. 
An example of how the foreground star contamination translate into magnitude
and color bias is presented in Fig. \ref{fgstarcontam}. It shows the
contamination of a galaxy, $K_s=12.5$, with a 12\arcsec\ diameter. 
To compute the probability of presence of a foreground star 
we use the starcount model described in \citet{Jar92} and \citet{JDH94},
with validation analysis presented in \citet{CBJC02}. The model also provides
us with the luminosity function at a given latitude and longitude. For the
worst case, at $l=90$ and $b=0.5$, the contamination could reach a few
percent of a magnitude in color (about 25\% probability that
$\Delta(J-K_s)>0.02$ mag). At this stage a new catalog is generated with
the statistical correction of the foreground star contamination bias. With
this correction, we expect color biases to be smaller than 1 to 2\%.

\subsubsection{Redshift contribution}
\begin{figure*}
	\centering
	\includegraphics[width=18.0cm]{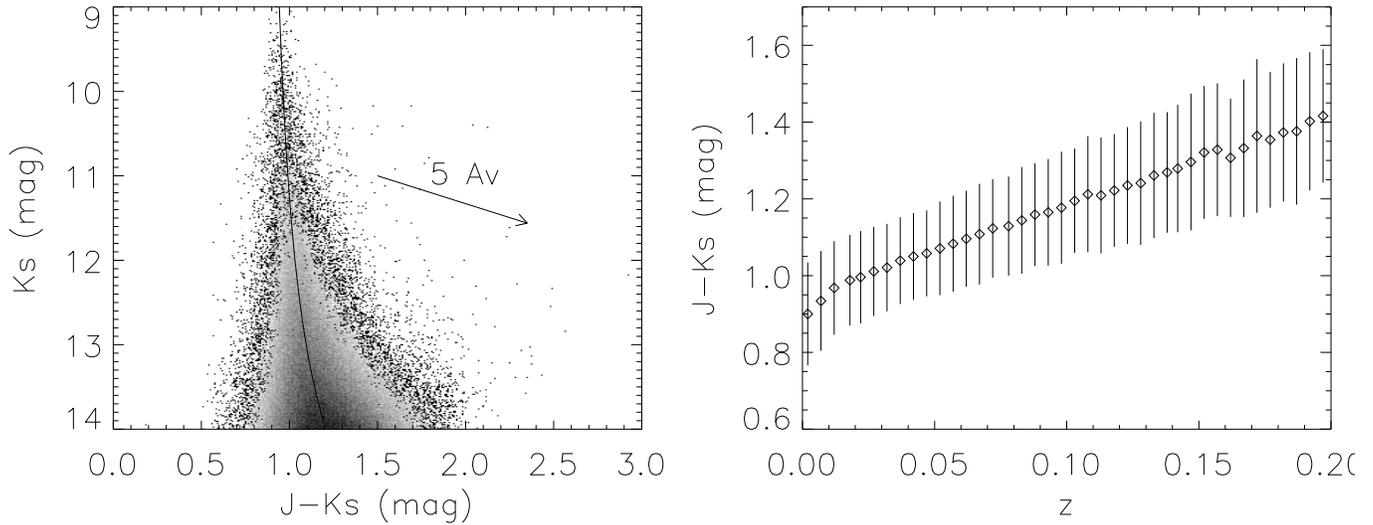}
	\caption{Left panel: color-magnitude diagram for galaxies non-reddened
	by the ISM ($I_{100 \mu {\rm m}}<1.5$ MJy sr$^{-1}$). The solid line
	indicates the magnitude-dependent reference for the extinction
	measurements. It is defined by the equation
	$J-K_s = (16.5-K_s)^{-1}+0.81$. Right panel: K-correction for $J-Ks$
	from \citet{Jar04}. The color variation in our galaxy sample is
	consistent with a maximum redshift of about 0.1.}
	\label{redshiftbias}
\end{figure*}
Given the 2MASS pixel size, galaxies as small as 10\arcsec\ are resolved as
extended sources, reaching $z \sim 0.1$ for the most luminous cluster
members \citep{Jar04}. At this redshift level, more $H$-band stellar light is
transfered into the $K_s$-band, reddening the $J-K_s$ galaxy color.
If individual redshifts for each source were known the method would be to
apply the {\em K-correction} to obtain the photometry that would have been
measured at $z=0$. However we are unable to make this correction as long as
we do not have the individual redshift knowledge.
It is still possible to correct the bias using the color-magnitude diagram
presented in Fig. \ref{redshiftbias}. The subsample selected to make this 
diagram contains only the object with a line of sight characterized by an
IRAS 100 $\mu$m surface brightness lower than 1.5 MJy sr$^{-1}$. The low
100 $\mu$m surface brightness ensures that the line of sight is free from
local ISM reddening. Assuming the fainter objects are the most distant observed,
the Fig. \ref{redshiftbias} trend is directly related to the
{\em K-correction}. The diagram shows a reddening of $J-K_s \sim 0.2$
mag between the bright and the faint end of the magnitude distribution which
is in agreement with the value expected for $z \sim 0.1$. The color excess
used to derive the extinction is therefore defined in the color-magnitude
plane as the distance along the reddening vector between each galaxy and the
reference line of equation $J-K_s = (16.5 - K_s)^{-1}+0.81$ (Fig.
\ref{redshiftbias}). The color excess is determined for each galaxy
following this definition. The statistical uncertainty on this
correction is negligible compared to the photometric uncertainty which
causes the scatter in Fig. \ref{redshiftbias}.

\subsubsection{Selection effect with the galactic latitude}
\begin{figure}
	\centering
	\includegraphics[width=8.0cm]{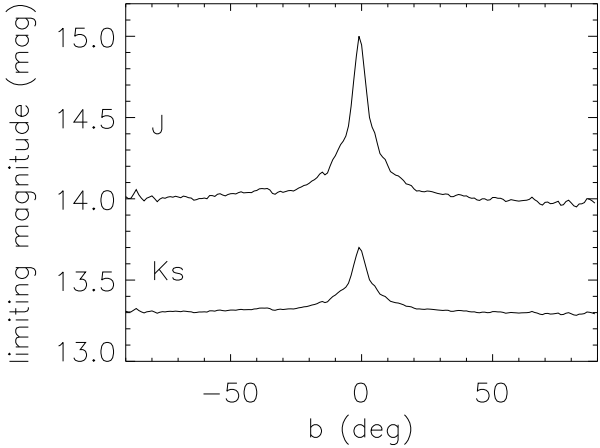}
	\caption{Latitude-dependent limiting magnitude used to correct the
	selection effect of galaxies reddened by the Milky Way disk dust. The
	curves represent the loss in sensitivity due to confusion noise.} 
	\label{maglim}
\end{figure}
Galaxies detected through the Milky Way disk suffer from more extinction than
high-latitude objects. Consequently the nature of the faint galaxies depends
on the latitude. For instance more high redshift objects are detected at high
latitude. This has been corrected in the redshift bias analysis, but one can
imagine other selection effect such as a low surface brightness galaxy
fraction or a face-on galaxy fraction as a function of the galactic latitude.
Another possible effect related to foreground extinction make the
fainter outer parts of galaxies to fall below the isophotal detection limit.
These outer parts are usually bluer because of the age and metallicity
gradients. The non-detection of the outer parts would make the galaxy color
to appear redder and the derived extinction to be overestimated. The
resulting bias depends on the photometric bands used. For $J-K_s$,
\citet{JCC+03} provide radial profile for about 500 large 2MASS galaxies
showing the color gradient is actually small (see their Fig. 15). The color
change is about 0.1 mag from the galaxy center to the outer disk, where most
of the change is happening well within the half-light radius. 
To reduce the selection effects we evaluate the average extinction as a
function of the latitude and we set a limiting magnitude which depends on the
galactic latitude. Figure \ref{maglim} shows that the magnitude cut is
about 1 mag brighter at $b > 30$ deg than it is at $b=0$ deg. The difference in
$K_s$ is smaller with about 0.4 mag. It is a strong constraint on the total
galaxy number which is reduced from $8.5 \times 10^5$ to $1.2 \times 10^5$
galaxies.

\subsection{Extinction mapping}\label{sect.mapping}
\begin{figure*}
	\centering
	\includegraphics[width=18cm]{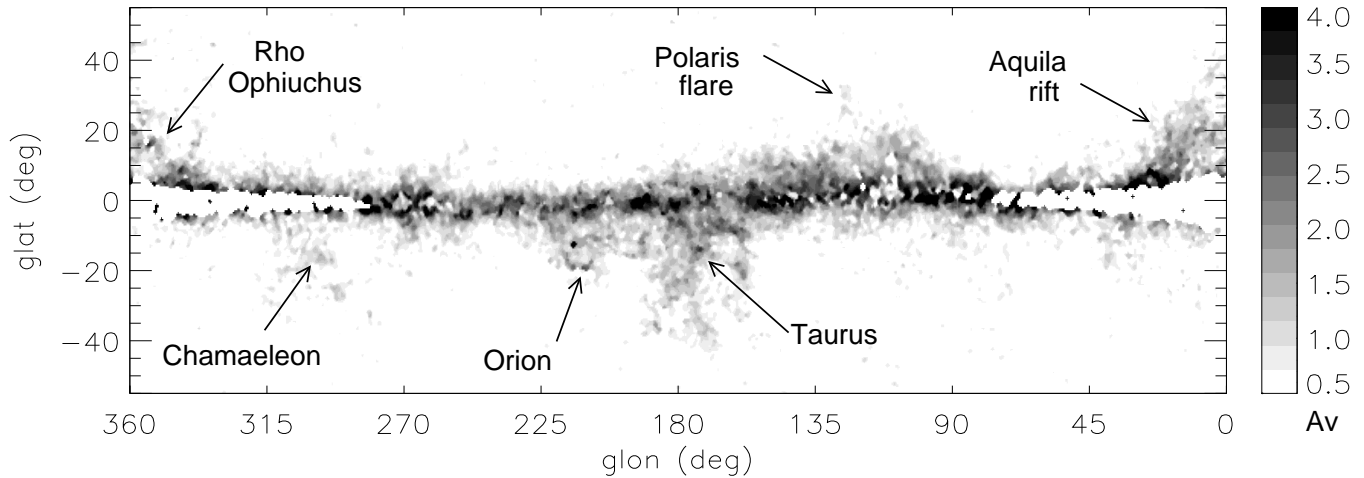}
	\caption{Extinction map derived from the median color in adaptive
	cells containing 5 galaxies each. The scale is labeled in visual
	magnitude units. The statistical uncertainty is 0.3 mag. As
	for Fig. \ref{galdistrib} no galaxy is detected near the galactic
	center due to high stellar density.}
	\label{avmap}
\end{figure*}
It is clear from Fig. \ref{galdistrib} that the galaxy density variations
follow the large scale structure of the universe \citep[see also][]{Jar04}.
The adaptive method described in \citet{CBJC02} to derive the extinction is
particularly well adapted in this case. For each position in the map the 
extinction is obtained from the median color excess of the 5 closest
neighbors. This technique optimizes the spatial resolution to the local
galaxy density. The result is presented in Fig. \ref{avmap}. The average
spatial resolution is 1 deg with variation from 30\arcmin\ in large scale
structures to about 4 deg for isolated galaxies. Such a low resolution map
should be viewed with caution since galaxies are not uniformly distributed in
space. As mentioned Sect. \ref{galcol} the galaxy color for a non-obscured
field is $J-K_s=1.0 \pm 0.1$. It translates into a statistical uncertainty
for the visual extinction of
$0.1 \times (A_J/A_V - A_{K_s}/A_V)^{-1} / \sqrt{5} = 0.3$ mag. However
the systematic error resulting from the spatial non-uniformity of the galaxy
distribution probably dominates the error. The extinction map obtained from
galaxy color excess is not equivalent to an average extinction at the given
resolution. The extinction map presented in Fig. \ref{avmap} should actually
be considered as a lower limit for the extinction since galaxies are
preferentially detected in low extinction areas, i.e. between dense clouds
that block the light coming from low surface brightness objects.

\section{Extinction and FIR dust opacity comparison}\label{results}
\subsection{Individual line of sight analysis}
Our goal is to compare the extinction measured from 2MASS galaxy color with
the dust optical depth in the FIR. IRAS data provides the 100 $\mu$m flux
density. The temperature is needed to translate the thermal dust emission to
optical depth. The work has been done by SFD98 using IRAS data modified to
take into account the temperature variations derived from lower resolution
DIRBE data. They have carried out the next step which is the conversion of the
optical depth into visual extinction. Since they performed a careful
calibration at high galactic latitude, we use their dust map.

The previous section pointed out that the extinction map presented in
Fig. \ref{avmap} is not equivalent to an average extinction at the cell
resolution. The situation is different for the FIR surface brightness maps
which provide the true arithmetic mean value at the beam resolution.
The direct comparison of the two maps is therefore not appropriate.
As an alternative to a map to map comparison we propose a strategy which
consists in comparing directly the individual galaxy reddening to the
corresponding SFD98 pixel extinction map. \citet{DAC+03} used a similar
strategy focusing on 20 early type galaxies for which they could
derive the extinction from their individual spectra. In the individual
comparison of galaxies with the SFD98 pixels, the spatial resolution problem
is reversed and single galaxies have the higher resolution with a typical
resolution of about 10\arcsec\ (the size of a galaxy) where the IRAS beam
size is only $\sim 5\arcmin$.

\subsubsection{Temperature bias}
The existence of several dust temperature components has been revealed by IRAS
through the analysis of the correlation between the 100 $\mu$m and 60 $\mu$m
surface brightnesses \citep{LCP91,ABMF94}. These results are actually based
on the observation of changes in the small grain abundances \citep{BFPH90}.
In \citet{LABP98} the temperature variations are directly measured using 
FIRAS spectra and DIRBE maps. They found a dust temperature of about 17.5 K
for the diffuse ISM and a colder component at around 15 K or less towards
molecular clouds.
\begin{figure}
	\centering
	\includegraphics[width=8.0cm]{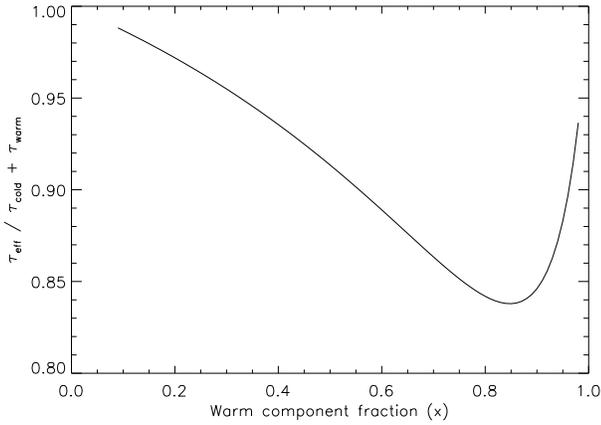}
	\caption{Figure from \citet{CBLS01} representing the effective to
	total optical depth ratio as a function of the warm component
	fraction. Cold and warm components are described by the modified
	Planck function $P=B_\nu(T) \times (\lambda/\lambda_0)^{-2}$ with
	$T_c=13.5$ K and $T_w=17.5$ K.} 
	\label{warmcold}
\end{figure}
In SFD98, the FIR optical depth is derived from the IRAS 100 $\mu$m
brightness using an effective temperature for a line of sight. It is not
equivalent to the sum of
the optical depths from different temperature components. The impact of this
point on the final optical depth is discussed in \citet{CBLS01} with a  model
assuming the 100 $\mu$m emission comes from a mixture of two distinct
components. Assuming a $\nu^2$ emissivity law, a single black body and a
combination of two black bodies are fitted to the emission. The result is
presented in Fig. \ref{warmcold}, which shows the effective optical depth is
always smaller than the sum of the optical depth for the two dust components.
When temperature changes along a line of sight, SFD98 are expected to 
underestimate the amount of interstellar material. On the other hand, the
temperature variations have absolutely no effect on the galaxy reddening.
There is a bias in the SFD98 optical depth determination which
depends on the fraction of warm dust component along a line of sight.
This fraction is hard to estimate and according to the modeling of Fig.
\ref{warmcold} a 5 to 15\% underestimation in the SFD98 extinction is
possible.

\subsubsection{ISM clumpiness bias}
\begin{figure}
	\centering
	\includegraphics[width=8.8cm]{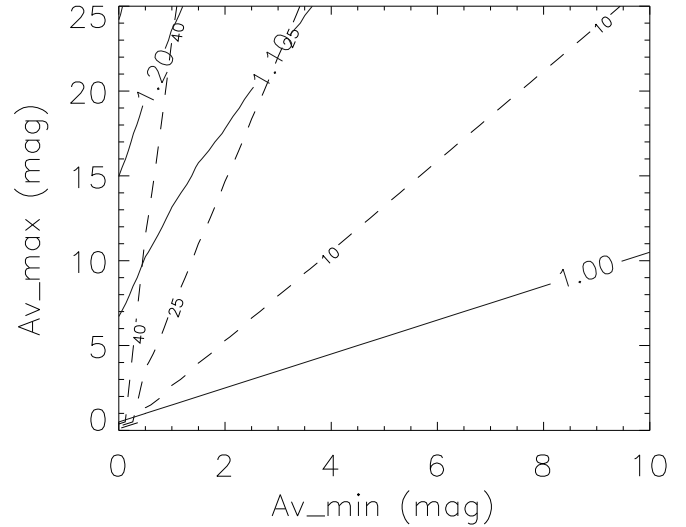}
	\caption{ISM clumpiness bias derived from a synthetic molecular
	clouds \citep{PJBN98}. The solid lines are the isocontours for
	$\overline{A_V}/A_V({\rm gal})$ and the dashed lines correspond to
	the $\sigma(A_V)/A_V$ ratio isocontours labeled in percentage. For
	$\sigma(A_V)/A_V \approx 25\%$, the bias in the extinction
	determination $(\overline{A_V}-A_V({\rm gal}))/A_V({\rm gal})$ is
	less than 10\%.}
	\label{ismclump}
\end{figure}
Section \ref{sect.mapping} focuses on the role of the galaxy distribution in
the extinction estimation and shows a non-linear effect results from the
non-uniform galaxy distribution. It is still relevant here because we 
compare each
galaxy to the corresponding IRAS pixel which is about 5\arcmin\ large. If
the interstellar medium is clumpy at the IRAS scale we expect a systematic
effect since galaxies will preferentially be detected in the less obscured
part of the pixel whereas the FIR extinction relies on the arithmetic mean
over the pixel. \citet{LAL99} and \cite{TBD97} have discussed the dust
clumpiness and they both conclude it is rather smooth compared to the gas. On
small scale they propose an upper limit for the $\sigma(A_V)/A_V$ ratio of
about 25\%. The $\sigma(A_V)/A_V$ ratio is a constraint on the dust
distribution but the scaling function is also needed to compute the bias.
\citet{PJBN98} generate synthetic molecular clouds from supersonic
turbulence. We use a sample of unreddened galaxies together with their
simulation to compute the bias. The method consists in applying the
extinction from the simulation and then to apply a magnitude cut which is
equivalent to the instrumental sensitivity limit. Then the ratio of the
arithmetic mean extinction to the median extinction from the remaining
galaxies gives the bias. We applied this method to various values for the
minimum and maximum extinction values in the simulated cloud; results are
presented in Fig. \ref{ismclump}. For each run the $\sigma(A_V)/A_V$ ratio
is computed and constrains the fluctuations expected from small scales. For 
$\sigma(A_V)/A_V \approx 25\%$ the galaxy color excess underestimates the
average extinction by about 10\% if the real extinction goes from 3 to 20 mag
within a single IRAS pixel. For lower extinctions, the systematic effect
decreases. It is safe to consider a 10\% bias as an upper limit.

\subsection{Results}
\begin{figure}
	\centering
	\includegraphics[width=8.8cm]{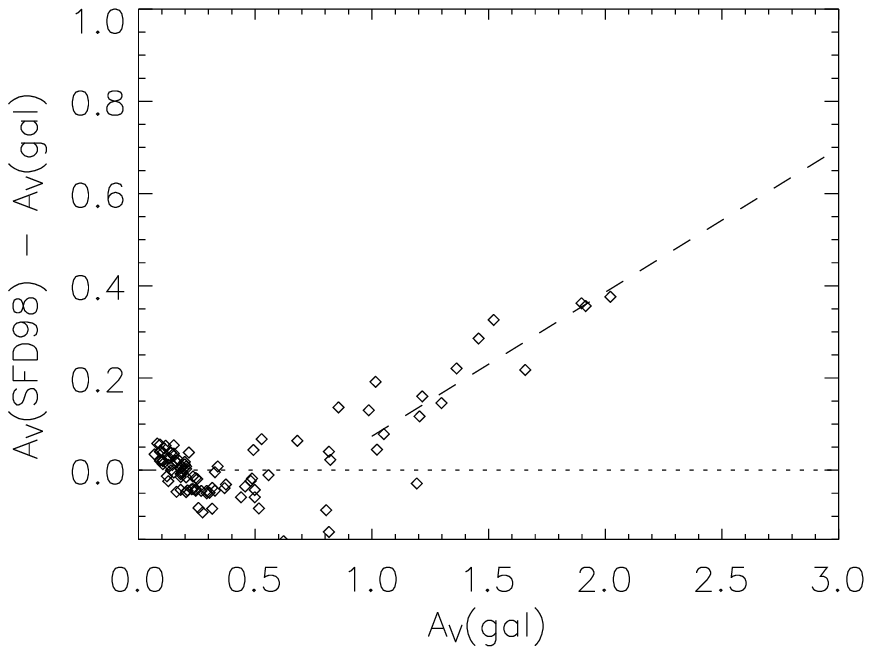}\\
	\vspace{0.5cm}
	\includegraphics[width=8.8cm]{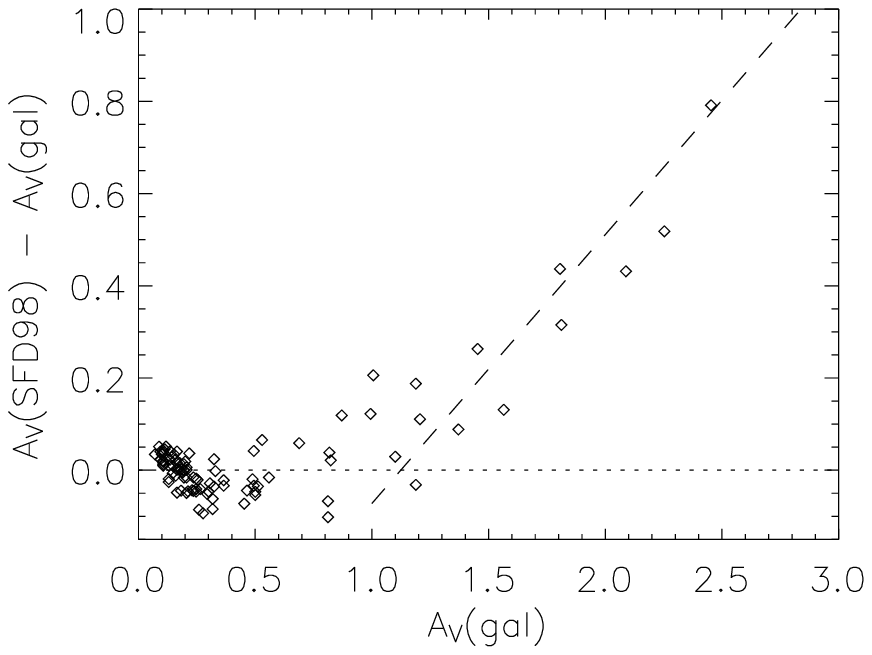}
	\caption{Upper panel: difference of extinction between SFD98 and
	2MASS galaxies versus the galaxy extinction, restricted to SFD98 pixels
	not contaminated or corrected for point sources (=covered by the PSCz
	catalog). The deviation for $A_V>1$ is fitted by the dashed
	line of equation $A_V({\rm FIR})=1.31 \times A_V({\rm gal})- 0.24$.
	Lower panel: same as upper panel but with all SFD98 pixels, including
	those for which the contamination cannot be assessed. The dashed
	line is $A_V({\rm FIR})=1.58 \times A_V({\rm gal})- 0.66$}
	\label{fig1}
\end{figure}
\begin{figure}
	\centering
	\includegraphics[width=8.8cm]{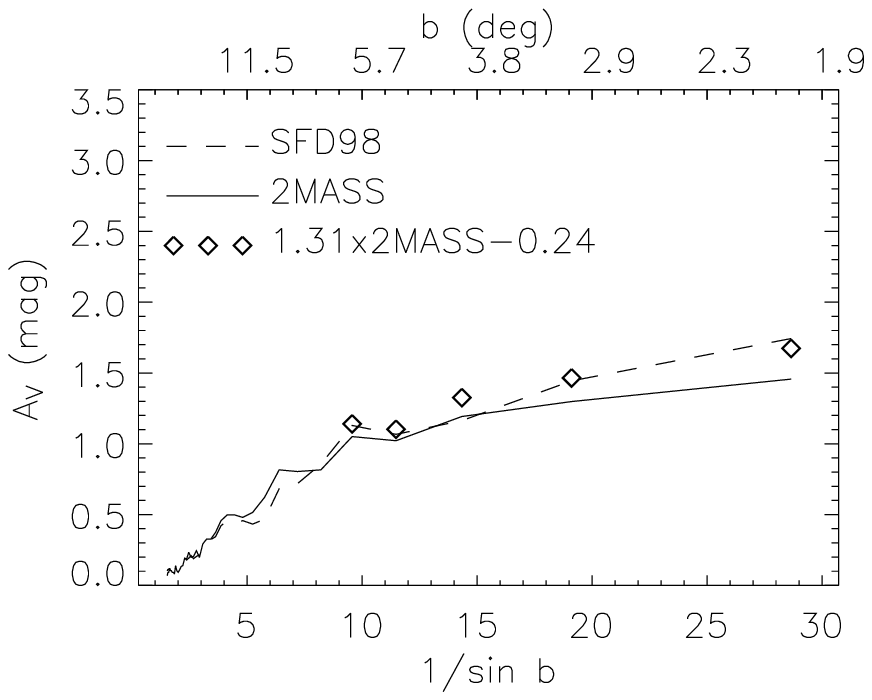}\\
	\vspace{0.5cm}
	\includegraphics[width=8.8cm]{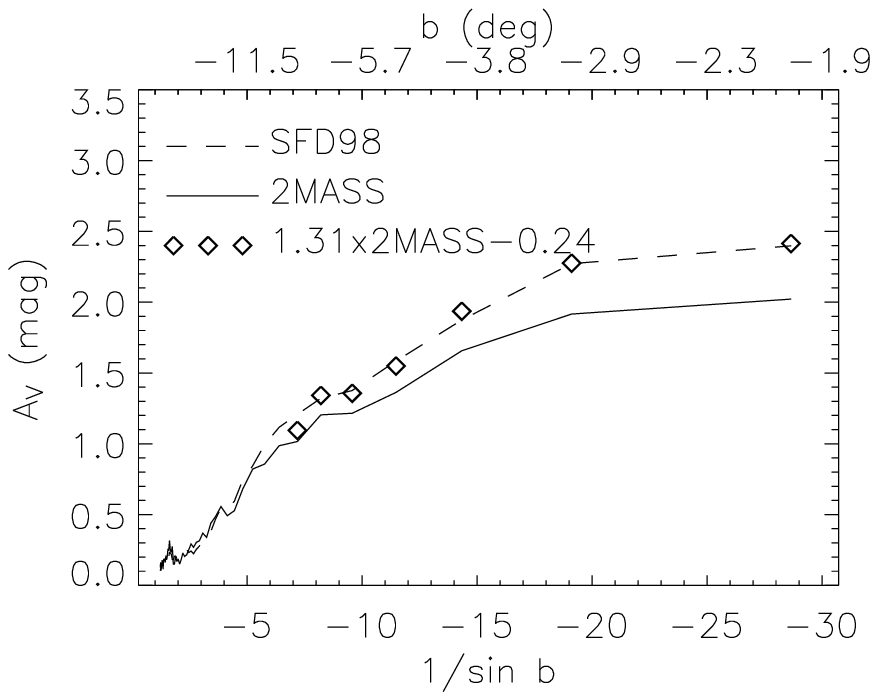}
	\caption{Extinction from 2MASS galaxies (solid line) and from SFD98
	(dashed line) versus the galactic latitude cosecant for the northern
	(upper panel) and the southern (lower panel) part of the anticenter
	hemisphere. The sample is restricted to SFD98 pixels not contaminated
	or corrected for point sources (=covered by the PSCz catalog).
	}
	\label{fig23}
\end{figure}
Our sample contains $1.2 \times 10^5$ galaxies corrected for all
biases described in Sect. \ref{bias123}. To compare the extinction obtained
for each galaxy with the FIR extinction map pixels we proceed as follows:
the two quantities are averaged on 1 deg wide strips in latitude that cover
the whole anticenter hemisphere longitude range (from 90 to 270 deg).
Almost all the spatial information is lost in this operation except the
latitude.
Point sources contaminate the IRAS fluxes. To derive their extinction
map \citet{SFD98} used the PSCz catalog \citep{SSM+00} to remove the point 
sources brighter than 0.6 Jy at 60 $\mu$m over most of the sky at $|b|>5$
deg and over parts of the sky at lower latitudes. The low-latitude sky is
critical in our analysis and we adopt a conservative approach. Among the
1149 low-latitude IRAS pixels ($0.5<b<1.5$ deg) containing a 2MASS galaxy
the contamination can be assessed for only for 259 (=23\%) of them. The
upper panel of Fig. \ref{fig1} shows the extinction difference between
SFD98 and 2MASS galaxies versus the galaxy extinction for pixels not
contaminated or corrected for point sources, i.e. in the area covered
by the PSCz catalog. For extinction greater than 1 mag a linear fit gives
$A_V({\rm FIR}) / A_V({\rm gal}) = 1.31 \pm 0.06$.
The lower panel is similar but includes all pixels, which means pixels not
contaminated or corrected for point sources {\em and} pixels for which the
contamination cannot be assessed. The ratio $A_V({\rm FIR}) / A_V({\rm gal})$
becomes 1.58. The discrepancy with the previous value is significant. We
chose to be conservative keeping only pixels covered by the PSCz catalog
which represent 23\% of our galaxy sample at $|b|<5$ deg and almost the whole
sky at $|b|>5$ deg. However pixels actually contaminated in the low-latitude
regions of the PSCz catalog represent only 7\% of the population. It suggests
we are probably underestimating the $A_V({\rm FIR}) / A_V({\rm gal})$ ratio
which would be in the range from 1.31 to 1.58.

Figure \ref{fig23} presents the same data in another form. It shows $A_V$
versus the latitude cosecant. The discrepancy appears more clearly in the
south because extinction is more significant in this area with the large
envelope surrounding the Taurus cloud.

\subsection{Discussion}
It is now understood that the SFD98 map overestimates the extinction at least
in some specific regions that have been carefully examined by various authors.
For instance \citet{AG99a} deduced an overestimation factor for SFD98
of 1.3 to 1.5 in the Taurus clouds, \citet{CFT+99} obtained
a 1.16 average factor towards 131 globular clusters at $|b|>2.5$ deg. A factor
of 2 was found by \citet{CBLS01} in the Polaris Flare which is a peculiar
region with an unexpected low temperature for a cirrus. In the galactic
center direction a discrepancy factor from 1.31 to 1.45 have been measured
using 2MASS data \citep{DSBB03,DSB02}.
In a recent paper \citet{DAC+03} studied the spectral properties of
20 early-type galaxies located at $|b|<25$ deg to derive the extinction
along the line of sight.
They found a ratio $A_V({\rm FIR})/A_V({\rm gal}) \approx 1.31$ for
$A_V>0.75$ and an agreement for lower values. Although their sample contains
only 20 objects they benefit from the high reliability of spectra.
Our analysis yields a very similar conclusion on a large scale with a
sample of $1.2 \times 10^5$ galaxies. \citet{DAC+03} and \citet{AG99a}
suggest the SFD98 map should be recalibrated, but they do not propose any
physical reason for this adjustment.

We support the idea of an enhanced emissivity of the dust in the
FIR due to fluffy grains \citep{CBLS01,SAB+03a}.
A grain growth through grain-grain coagulation or accretion of gas species is
expected in dense cold molecular clouds \citep{Dra85a} and leads to porous
grains. \citet{Dwe97} studied the fluffy grain optical properties and showed
the emissivity in the FIR is increased for porous grains. If this
effect applies to the present analysis it means the grain coagulation starts
at extinction close to 1 mag. We would expect variations from cloud
to cloud depending on their star forming activity, the local interstellar
radiation field and their geometry. The minimum extinction value at which
the FIR to extinction ratio changes is actually not the correct quantity
to check since it relies on the column density rather than the 3D volume 
density. Several low density clouds on the same line of sight and a single
dense cloud may have the same column density but their
$A_V({\rm FIR})/A_V({\rm gal})$ would be very different.
However, Fig. \ref{fig23} points out the extinction values beyond the threshold
for the enhanced emissivity are generally restricted to low-latitude regions
with $|b|<8$ deg. Our interpretation for the grain aggregate formation leads
to the conclusion that higher-density clouds are preferentially located in
the plane, which is in total agreement with our knowledge about the ISM
distribution in the Galaxy.
A comparative analysis for various environment would confirm our hypothesis
if clouds were proved to be different in their FIR to extinction correlation.
Unfortunately the spatial information in our analysis, the galactic latitude,
does not allow us to conclude on spatial variation of the
$A_V({\rm FIR})/A_V({\rm gal})$ ratio.

\section{Conclusion}\label{conclusion}
The 2MASS galaxy colors appears to be a very interesting extinction estimator
to compare with FIR extinction. Both extinction estimators are sensitive to
the absorption for the whole line of sight, but they are totally independent.
{\em We have investigated the galactic anticenter hemisphere and found
$A_V({\rm FIR}) / A_V({\rm gal}) = 1.31 \pm 0.06$ for the SFD98 extinction map
restricted to pixels corrected for point source contamination}.
We correct for several biases that would affect the correlation analysis. 
The foreground star contamination estimated with the stellar density number
and the galaxy size has been found negligible with about 1-2\% uncertainty.
Redshifted galaxy photometry has been corrected using a color-magnitude
diagram. The correction is consistent with the expected {\em K-correction}
for $z \la 0.1$.
The presence of several dust temperature components on a line sight implies a
systematic effect in the FIR extinction map. This effect yields a 5 to 15\%
underestimation of the FIR extinction which reduces the discrepancy between
FIR and galaxy color extinction. The difference in the spatial resolution of
the extinction value (6.1\arcmin\ for SFD98 and $\sim 10\arcsec$ for 2MASS
galaxies) roughly compensates the temperature systematics because of the ISM
clumpiness.

In this paper we have generalized previous studies that had already concluded
the SFD98 map overestimates the extinction for some specific directions to
half of the sky. At large scale the discrepancy appears for $A_V \ga 1$,
suggesting the whole galactic disk is affected. We attribute our result to
the presence of fluffy/composite grains which have an enhanced emissivity in
the FIR. Our large scale study suggests they are more common that previously
thought since they would be formed even at low extinction and not only in
dense cold clouds. More analysis are needed to confirm this point especially
to recover more spatial information. Complementary FIR and submm data from
Spitzer, Herschel and Planck are expected to solve this problem by providing
dust spectra at long wavelengths.

\begin{acknowledgements}
We acknowledge P. Padoan for providing us with the synthetic interstellar
cloud data.
This publication makes use of data products from the Two Micron All Sky Survey,
which is a joint project of the University of Massachusetts and the Infrared
Processing and Analysis Center/California Institute of Technology, funded by
the National Aeronautics and Space Administration and the National Science
Foundation.
\end{acknowledgements}

%\bibliographystyle{aa}
%\bibliography{aamnem99,biblio}

\end{document}